# HARDWARE IMPLEMENTATION OF ALGORITHM FOR CRYPTANALYSIS

Harshali Zodpe[1] , Prakash Wani[2] , Rakesh Mehta[3]

[1]Department of Electronics and Telecommunication Engineering, Maharashtra Institute of Technology, Pune, India
harshaliz1982@gmail.com

[2]Department of Electronics and Telecommunication Engineering, College of Engineering, Pune, India
pwwani@gmail.com

[3]Mechatronics Test Equipments (I) Pvt. Ltd., Pune, India
rakesh.mehta57@gmail.com

## ABSTRACT

*Cryptanalysis of block ciphers involves massive computations which are independent of each other and can be instantiated simultaneously so that the solution space is explored at a faster rate. With the advent of low cost Field Programmable Gate Arrays (FPGA's), building special purpose hardware for computationally intensive applications has now become possible. For this the Data Encryption Standard (DES) is used as a proof of concept. This paper presents the design for Hardware implementation of DES cryptanalysis on FPGA using exhaustive key search. Two architectures viz. Rolled and Unrolled DES architecture are compared and based on experimental result the Rolled architecture is implemented on FPGA. The aim of this work is to make cryptanalysis faster and better.*

## 1. INTRODUCTION

The extensive use of computers and electronic data storage and transmission with the exponential growth of internet has generated strong demand for a robust security mechanism, and they in turn depend critically on cryptographic protection. The computer power is also increasing fast with the change in technology and it is important to assess the strength of deployed cryptographic algorithm.

Cryptanalysis is the science of revealing the hidden data of a cryptographic algorithm or system, either to get at the data for the own sake or to test the strength of the cryptographic algorithm being used. Cryptanalysis of ciphers (encrypted information) usually involves massive and parallel computations. The security parameter (in particular the key length) of almost all practical crypto algorithms is chosen such that attacks with conventional computers are computationally infeasible. Such parallel functionality can be realized by special purpose hardware blocks that can be operated simultaneously, improving the time complexity of the overall computations. But the high non-recurring engineering cost for Application Specific Integrated Circuits (ASIC's) had put most projects for building special purpose hardware for cryptanalysis out of the reach for commercial or research institutions.

However, Reconfigurable Computing offers advantages over traditional software and hardware implementations of computationally intensive algorithms. Reconfigurable computing is based on using low cost FPGA's which can be configured after fabrication to take advantage of a hardware design but still maintain the flexibility of software. Thus Cryptanalytic hardware has now become a possibility outside government agencies.

Depending on what information the adversary can obtain the different attack scenarios can be distinguished as Ciphertext-Only attack, Known-Plaintext attack, Chosen-Plaintext attack, Chosen-ciphertext attack, Adaptive Chosen-Plaintext/Ciphertext attack as published in [1].

This paper presents a FPGA based hardware design for cryptanalysis of DES based on known-plaintext attack using brute force technique. The DES algorithm is chosen for implementation as it is basic cryptographic algorithm and is used by academicians as a test unit for experimentation. The idea is to create multiple instances of the key search engine in an FPGA chip to make the cryptanalysis process faster and provide a better cost performance ratio.

## 2. HISTORICAL BACKGROUND

Cryptanalysis has a historical background. Cryptanalysis is practiced by a broad range of organizations to test the strength of the cryptosystem being used. The first exhaustive DES key search machine estimation was proposed by Diffie and Hellman in 1977 [2]. It contained $10^6$ chips, with an estimated cost of US$ 20 million and a 12 hour expected search time. After few years, a first detailed hardware design description for a brute force attack was presented by Michael Wiener at CRYPTO'93 [3]. It was estimated that the machine could be built for less than a million US dollars. The proposed machine consisted of 57, 000 DES chips that could recover a key every three and half hours. In 1997, a detailed cost estimate for three different approaches for DES key search, distributed computing, FPGAs and custom ASIC designs, was presented by M. Blaze [4]. In 1998, the Electronic Frontier Foundation (EFF) finally built a DES hardware cracker called Deep Crack which could perform an exhaustive key search within 56 hours [5]. Their DES cracker consisted of 1, 536 custom designed ASIC chips at the material cost of around US$ 250,000 and could search 88 billion keys per second. In 2006, the Cost Optimal Parallel Code Breaker (COPACOBANA) for DES brute-force attack was built for less than US$ 10,000 [6]. COPACOBANA hosts 120 low-cost FPGAs and is the latest developed cryptanalysis hardware capable of breaking DES in less than one week on average. With the emergence of newer and powerful devices, continuous efforts are being made to make cryptanalysis faster and better.

## 3. DESCRIPTION OF DES ALGORITHM

The Data Encryption Standard (DES) is one of the first commercially developed ciphers. DES was published as a U.S. Federal Information Processing Standard, FIPS-46 [7]. DES is a block cipher operating on 64-bit data blocks. The encryption transformation depends on a 56-bit secret key and consists of sixteen rounds surrounded by two permutation layers. The decryption process is the same as encryption, except the order of the round keys is reversed as compared to the encryption process as shown in Figure 1.

As seen in Figure1, the first block is initial permutation. It changes the order of the input bits according to a fixed permutation. The result is then divided into two equal halves of 32-bit each i.e. $L_i$ and $R_i$ where 'i' denotes the round number. In each round, the previous word $R_i$ is fed to a round function 'f' and the result is then XORed to the previous word $L_i$. Both the words are then swapped and the algorithm proceeds to the next iteration. The round function 'f' is key dependent and involves the following steps. The first step is expansion. Here the 32-bit input word is expanded to 48-bits by duplicating and reordering the right half of the bits. This input is then

XORed with the required key depending on the round number in the second step. In the third step, the 48-bits output from previous step is split into eight 6-bits words which are substituted in eight parallel 6x4 bit S-boxes. This substitution increases the strength of the cryptosystem. The last step is permutation. In this step the 32-bit output from previous step is reordered according to a fixed permutation.

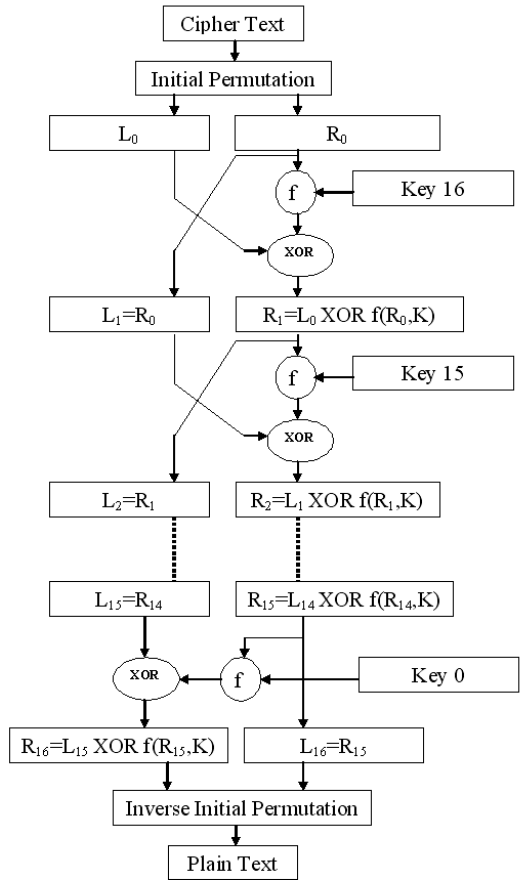

Figure 1- DES Decryption Process

The subkeys (keys generated from the given input key) required for the 16-rounds are calculated by the key schedule algorithm. Figure 2 shows the key schedule algorithm. The block PC1 of the key schedule algorithm discards the 8 parity bits from the input 64-bit key and divides the resultant 56-bits into two halves of 28-bit each i.e., $C_i$ and $D_i$. These are cyclically rotated over one position to the left after rounds 1, 2, 9, 16 and over two positions after all other rounds. The round keys are then constructed by repeatedly extracting 48-bits from $C_i$ and $D_i$ at 48 fixed positions determined by block PC2. Thus the keys generated are Key1 to Key16 for the 16-rounds. For decryption process the subkeys have to be applied in reverse order. Hence SRL16 block is used to reverse the order of subkeys. The SRL16 is an alternative mode for the look-up tables where they are used as 16-bit shift registers. These shift registers enable delay or latency compensation.

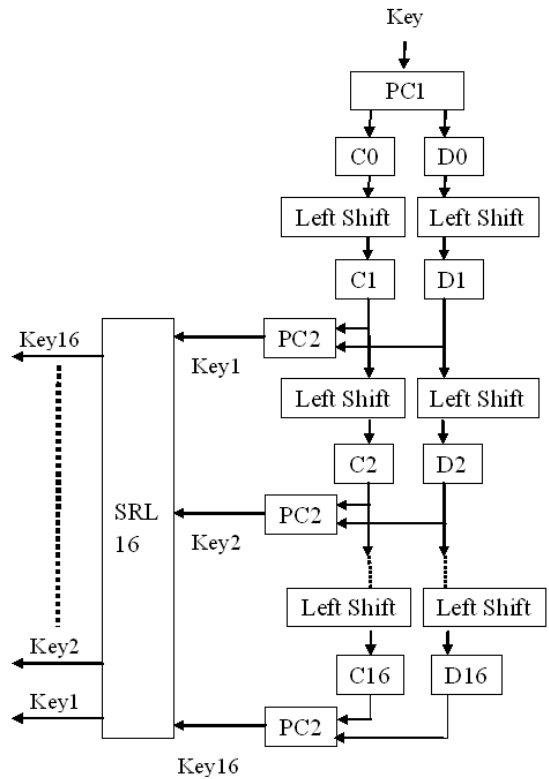

Figure 2 – Key Schedule Algorithm

## 4. ARCHITECTURAL CONSIDERATIONS FOR HARDWARE IMPLEMENTATION

The design is implemented considering architecture options for speed area optimization. For cryptanalysis the ciphertext is deciphered under each key and the result is compared with the known plaintext. If they are equal, then it is possible that the key tried is the correct key. Thus for cryptanalysis first decryption has to be performed with each key. The decryption takes 16-rounds to decipher the ciphertext. DES algorithm contains an iterative structure. Data is passed through the same set of steps called 'round' 16 times, each time with a different sub-key from the key transformation. The first architecture is implemented using Rolled technique where only one round is design and using a multiplexer, the same round is used 16 times with different sub-key for each round as shown in Figure 3. This architecture is area efficient but with lesser throughput.

As seen from Figure 3, the incoming data is passed through the initial permutation. The output from IP block is then passed 16-times through the round along with the 16 sub-keys generated by the 'subkey gen' block. In order to loop the output back to the input multiplexer is used. The multiplexer switches the input's of the data from the previous round and the new input data. The second architecture that is implemented is the Unrolled architecture as shown in Figure 4. Unrolling an iterative loop increases throughput. Here 16 instances are created for 16-rounds. In this design better throughput can be achieved at the cost of increased area utilization. The throughput can be further improved by using pipelining.

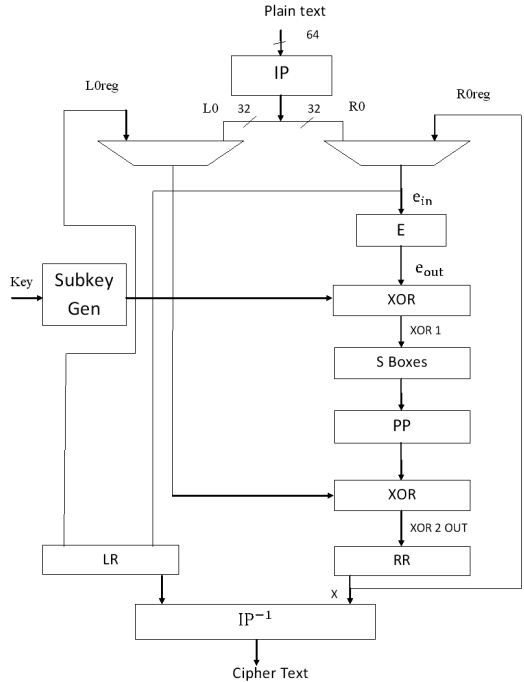

Figure 3 - DES Decryption using Rolled architecture

For pipelining registers are inserted between each step. In a pipelined design the new data can begin processing before the prior data has finished processing. However, there exists a degree of pipelining that maximizes the throughput per unit area.

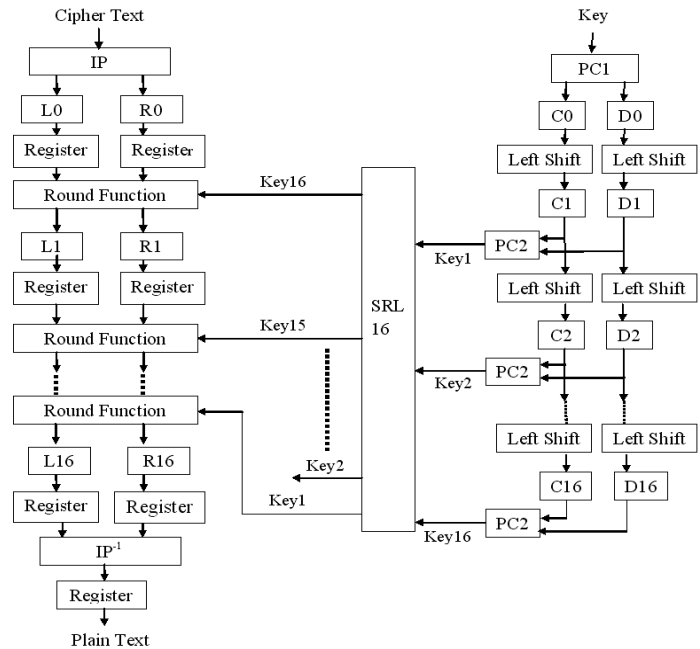

Figure 4 - DES Decryption using Unrolled architecture with pipelining

## 5. Cryptanalysis of DES

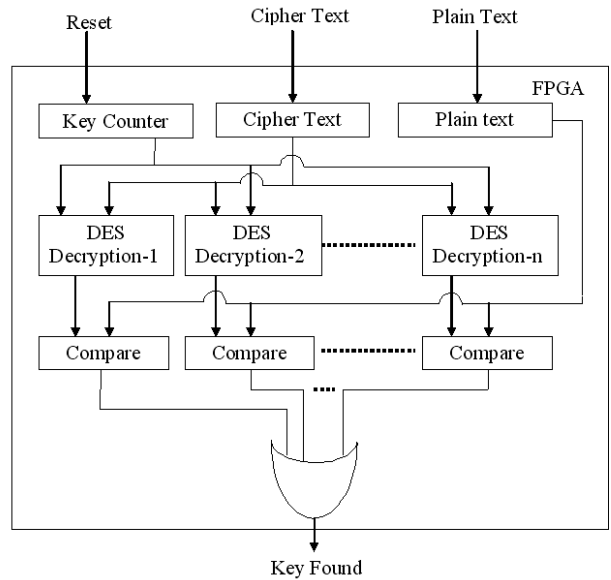

Figure 5 – Block Diagram for Cryptanalysis of DES

This paper presents a FPGA based hardware design for cryptanalysis of DES based on known-plaintext attack using brute force technique. A known-plaintext attack requires the adversary to have access to (part of) the plaintext corresponding to the captured ciphertext blocks. In the proposed design using brute force technique, the captured ciphertext block is decrypted with all possible keys and the resultant plaintext is compared with the know-plaintext. As shown in Figure 5, the DES Decryption block decrypts the ciphertext and the resultant plaintext is compared with the known-plaintext. The key for which the resultant plaintext matches with the known plaintext is considered to be the correct key.

Each DES Decryption block decrypts the ciphertext with different key. Thus with 'n' nos. of DES Decryption blocks 'n' different keys can be searched in one clock cycle thereby reducing the time required for key search by a factor of 'n'. The no. of DES Decryption blocks 'n' depends on the available logic resources in an FPGA and the logic utilization of one DES Decryption block.

## 6. IMPLEMENTATION OF DES CRYPTANALYSIS

The aim of the hardware is the probable key search to be accomplished by partitioning the solution space on the FPGA chip and instantiating multiple instances of the design in parallel. As the solution space is independent and inter-process communication is hardly required, the key search time can be reduced by 'n' fold for 'n' instances of the design in single a FPGA chip.

We have implemented the design in Virtex-4 FPGA chip (xc4vlx100-12ff1148) using Xilinx 13.1 development platform. The design is implemented using both Rolled and Unrolled architecture. After synthesis of the design incorporating  key search engines in a single FPGA along with the additional logic required for finding the correct key, the device utilization and throughput for both the architectures are as shown below.

### 6.1. Using Rolled Architecture

The usage of 1,662 slices, 892 slice flip-flops (FF) and 3,187 Look up Tables (LUTs) is reported by the tool. (3% slices, 0% slice FF and 3% LUT utilization of the xc4vlx100-12ff1148 device respectively). The throughput achieved for the design is 0.817GBPS with maximum frequency of 230.063MHz.

### 6.2. Using Unrolled Architecture

The usage of 9,019 slices, 7,614 slice flip-flops and 16,954 LUTs is reported by the tool. (18% slices, 7% slice FF and 17% LUT utilization of the xc4vlx100-12ff1148 device respectively). The throughput achieved for the design is 0.874GBPS with maximum frequency of 245.874MHz.

From the above experimental results, the device utilization for design using Rolled architecture is considerably less as compared to design using Unrolled architecture. Hence the Rolled architecture was selected for implementation. 256 key search instances for the Rolled architecture could be fit in a single FPGA with a clock frequency of 323.515MHz. With the clock frequency achieved, the entire key space can be searched in approximately 5 days. A comparative summary is shown in Table 1 with the results obtained in [6].

| | Ours | [6] |
|---|---|---|
| No. of FPGA's | 01 | 120 |
| No. of instances per FPGA | 256 | 04 |
| Device utilization per FPGA | 25% Slice registers and 96% Slice LUT's | --- |
| Operating Frequency per FPGA | 323.515MHz | 100MHz |
| Time for Exhaustive Key search | Approximately 5 days | Approx. 8.7 days |

Table 1. Implementation Results

## 7. RESULTS

### 7.1. Using Rolled Architecture

I/P- plaintext<=x"12cf4d587bf4eb08";

I/P- ciphertext<=x"b6060c26730925bc";

O/P- Key<= x"00000000000001";

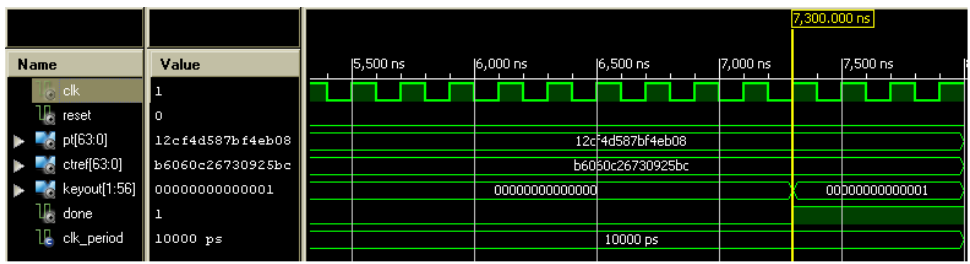

Figure 6. Simulation result for key= x"00000000000001"

13

I/P- plaintext<=x"12cf4d587bf4eb08";

I/P- ciphertext<=x" 91dbf8a0e3f63324";

O/P- Key<= x"f0000000000011";

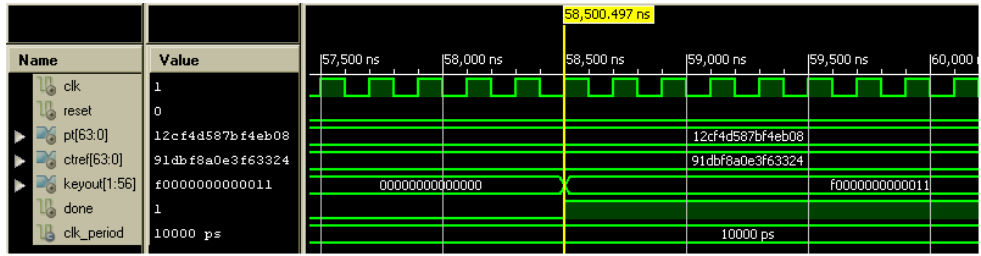

Figure 7 - Simulation result for key= x"f0000000000011"

Referring to Figures 6-7, using the Rolled architecture with the given plaintext-cipher text pair, the correct key x"00000000000001" and x"f0000000000011" is obtained at 7,300 ns and 58,500 ns respectively. Using single instance of the key search engine would take months to search the correct key x"f0000000000011". By running four instances in parallel the search time has considerably reduced.

## 8. CONCLUSION

This work presents the design for cryptanalysis of DES algorithm based on Rolled architecture using FPGA. Based on the experimental results it is found that the Rolled architecture requires less area and can search the entire solution space in less time as compared to the Unrolled architecture. In this work we could fit 256 instances of key search in a single FPGA, such that the exhaustive key search on DES Algorithm can be done in approximately 5 days. Expanding the concept for AES Algorithm to fit maximum instances of the key search engine in a single FPGA and implementing the design in FPGA is the future work of this paper.

### ACKNOWLEDGEMENTS

We would like to thank the entire team from Mechatronics Test Equipments (I) Pvt Ltd, Pune for their valuable guidance and support.

## Authors

Ms.Harshali Zodpe completed her B.E. in Electronics and Communication Engineering from Visvesvaraya National Institute of Technology, Nagpur, M.E in VLSI Design from Ramdeo-baba Kamla Nehru College of Engineering, Nagpur University, Nagpur. Her areas of interests are VLSI, cryptography & Embedded Systems. Currently she is pursuing her Ph.D from College of Engineering, Pune.

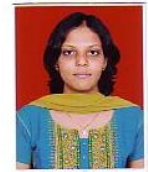

Dr. P.W.Wani has received his M.E. and Ph.D in Electroncis and Telecommunication from College of Engineering, Pune. His areas of interest are VLSI, Microelectronics and parallel Processing. He was the dean academics of Pune University. Currently he is working as a professor in the department of Electroncis and Telecommunication from College of Engineering, Pune.

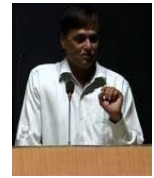

Mr. Rakesh Mehta received his B.E from College of Engineering Pune and M.Tech from IIT, Madras. He has been an entrepreneur for the past 25 years and his areas of expertise are Digital Design and FPGA based High Speed designs. Currently he is the Director, Mechatronics Test Equipment and Bitmapper Integration Technologies, Pune.

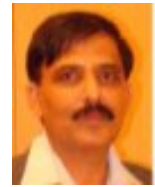